\documentclass[12pt]{article}
\usepackage{amsmath, amssymb, amstext, amsthm}
\usepackage{graphicx}
\usepackage{epstopdf}
\usepackage{cite}
\usepackage{color}
\usepackage{hyperref}
\usepackage{braket}
\usepackage{slashed}
\usepackage{cancel}
\usepackage{multirow}
\usepackage{float}
\usepackage{bm}
\usepackage{lineno}
\usepackage{subfigure}

\begin{document}
	
\begin{titlepage}
	\begin{center}
		
		\vspace*{15mm}
		
		\vspace*{10mm}
		
		{\Large{ \bf Constraints on a Light Singlet Scalar from Combined Exotic Higgs Decays}}
		\vspace{1cm}
		
		{\small \bf {F. Azari\footnote{{fh.azari20@gmail.com}}, and M. Haghighat\footnote{Corresponding author:\,{m.haghighat@shirazu.ac.ir\, ; \, mhaghighat44@gmail.com}}} }
		
		\vspace*{0.5cm}
		
		{\it \small Physics Department, College of Sciences, Shiraz University 71454, Shiraz, Iran}\\
		
	\end{center}

\begin{abstract}
We investigate the phenomenology of the Standard Model extended by a real gauge-singlet scalar field, focusing on exotic Higgs decay channels. For a light scalar mass in the range \(0 < m_{\phi} < 40\) GeV, the Higgs boson can decay to both two and three scalar final states. We derive analytical expressions for these decay rates and impose a global constraint on the model parameters by requiring that their sum does not exceed the total Standard Model Higgs boson decay width. This requirement translates into a fourth-order inequality with respect to the singlet vacuum expectation value, \(v_{\phi}\). We demonstrate that satisfying this inequality imposes an upper bound of \(\cos \theta < 0.12 - 0.13\) across the entire mass range, providing a complementary constraint to existing direct search limits. Utilizing stronger independent constraints on the mixing (e.g., \(\cos \theta < 0.1\)), we then predict upper bounds on the individual exotic decay rates as a function of \(m_{\phi}\) as \(\Gamma_{h \rightarrow \phi \phi} < 0.06\) MeV and \(\Gamma_{h \rightarrow \phi \phi \phi} < 5 \times 10^{- 6}\) MeV, respectively.
\end{abstract}
\end{titlepage}	
\newpage

\section{Introduction}

The Standard Model (SM) of particle physics stands as one of the most successful scientific theories, providing a framework that accurately predicts a vast array of experimental results \cite{Glashow1961, Weinberg1967, Salam1968}. Despite its triumphs, the SM is incomplete and cannot explain fundamental phenomena such as the nature of dark matter and dark energy. A cornerstone of the SM's success was the prediction and subsequent discovery of the Higgs boson at the LHC \cite{Higgs1964, Englert1964, Guralnik1964, ATLAS2012, CMS2012}. This discovery opened a new window for probing the Higgs sector and testing its possible extensions, which are a primary path for exploring physics Beyond the SM (BSM).

Light scalar singlets appear naturally in several well-motivated BSM scenarios. They can serve as Higgs portal dark matter candidates, where the scalar acts as a thermal relic \cite{Silveira1985, McDonald1994}. They may arise as pseudo-Goldstone bosons from approximate global symmetry breaking (axion-like particles) or as mediators between the SM and a hidden sector. Furthermore, light scalars can modify the Higgs potential and facilitate a strong first-order electroweak phase transition, which is essential for electroweak baryogenesis \cite{Profumo2015}. For comprehensive reviews of singlet scalar extensions and their phenomenology, see Ref. \cite{PhysRept2012}.

Among the myriad of BSM scenarios, extensions of the scalar sector are particularly well-motivated. These range from minimal additions, such as a single real gauge-singlet scalar field, to more complex structures like Two-Higgs-Doublet Models \cite{Veltman1989, Silveira1985, McDonald1994, Schabinger2005, Patt2006, OConnell2007, Barger2008, Profumo2007, Baek2012, Robens2015, Godunov2016, Falkowski2015, Bertolini2012, Englert2013}. The simplest case---adding a scalar singlet with a non-zero vacuum expectation value---is especially compelling \cite{Veltman1989}. It can couple to the SM Higgs field at the renormalizable level \cite{Silveira1985} through a portal interaction:

\begin{equation}
V_{h\phi} = \lambda_{3}H^{\dagger}H\Phi^{2}. \label{eq:portal}
\end{equation}

This coupling induces a mixing between the SM Higgs \((H)\) and the new scalar \((\phi)\), introducing two new fundamental parameters: the mixing angle \((\theta)\) and the mass of the new scalar particle \((m_{\phi})\). This simple extension has been extensively studied \cite{Profumo2015, Chen2015, Costa2015, Pruna2013, CMS2013, No2014}, with significant effort dedicated to constraining the mass and mixing angle, as summarized in Tables 1 and 2. These constraints often arise from Higgs coupling measurements, electroweak precision observables, and direct searches, and have also been considered in the context of vacuum stability \cite{Ema2018, Lebedev2013, Alekhin2012, EliasMiro2012, Lebedev2012} and theoretical self-consistency \cite{Bezrukov2012, Lebedev2013b, Buttazzo2013}.

Extensive previous work has established constraints on the singlet parameter space. For heavy scalars \((m_{\phi} > 50 \mathrm{GeV})\), LHC measurements typically allow \(\cos \theta < 0.3 - 0.4\) \cite{Arcadi2019, Profumo2015}, while future lepton colliders could probe \(\cos \theta < 0.09 - 0.19\) \cite{Profumo2015}. For light scalars \((m_{\phi} < 40 \mathrm{GeV})\), LEP and LHC searches constrain \(\cos \theta < 0.07 - 0.17\) depending on mass \cite{Robens2015}, with electroweak precision tests allowing \(\cos \theta < 0.4\) \cite{LopezVal2014}. However, these studies often focus on individual channels, and as a result, no comprehensive constraint exists from combining exotic Higgs decay channels \(h \to \phi \phi\) and \(h \to \phi \phi \phi\).

One particularly underexplored avenue for constraining light singlets \((m_{\phi} < m_{H} / 2)\) is through exotic Higgs decays. While previous studies frequently consider individual decay channels like \(H \to \phi \phi\) in isolation, the combined contribution of all kinematically accessible exotic decays provides a more global and powerful constraint. Specifically, for a very light scalar satisfying \(3m_{\phi} < m_{H}\), the Higgs can decay not only to two scalars \((H \to \phi \phi)\) but also to three \((H \to \phi \phi \phi)\) \cite{Helmboldt2017}. However, the combined contribution of both decays has not been used to derive constraints on the model parameters. Imposing a limit on their sum provides a more comprehensive and stringent bound than considering either channel in isolation.

In this work, we demonstrate the significant constraining power of a combined, multichannel analysis. We calculate the rates for both the \(H \to \phi \phi\) and \(H \to \phi \phi \phi\) decays for a light scalar field \((0< m_{\phi}< 40 \mathrm{GeV})\). Crucially, we impose a constraint by requiring that their sum does not exceed the experimentally permitted budget for BSM decays, which we estimate from the residual width not accounted for by the dominant SM decay channels \((b\bar{b}, W^{-}W^{+}, gg, \tau^{+}\tau^{-}, c\bar{c}, ZZ)\). This approach yields a stringent fourth-order inequality in the singlet vacuum expectation value \((v_{\phi})\), leading to novel limits on the mixing angle \(\theta\). By further incorporating external constraints on the mixing, we derive updated upper bounds on the individual exotic branching ratios across the entire viable mass range.

This paper is organized as follows: In Section II, we briefly introduce the formalism of the SM extended by an extra singlet scalar field, clarifying the notation and parameter counting. In Section III, we present our calculation of the exotic Higgs decay rates to two and three scalars, including complete phase space integrals, and derive the new constraints on the model parameters. Finally, we summarize our results and present our conclusions in Section IV.

\begin{table}[h]
\centering
\caption{Limits on the mixing angle and mass of the new scalar particle from various experiments for heavier mass ranges.}
\begin{tabular}{|l|c|c|l|}
\hline
\textbf{Production \& decay channels} & \textbf{\(m_{\phi}\) [GeV]} & \textbf{\(\cos \theta\)} & \textbf{Experiment} \\
\hline
Production \& decay channels & - & $< 0.33$ & LHC \cite{Arcadi2019} \\
Production channels & - & $< 0.4$ & LHC \cite{Profumo2015} \\
Higgs measurements & $50 < m < 250$ & $< 0.3$ & HL-LHC \cite{Profumo2015} \\
Higgs measurements & $50 < m < 250$ & $< 0.19$ & ILC-I \cite{Profumo2015} \\
Higgs measurements & $50 < m < 250$ & $< 0.13$ & ILC-3 \cite{Profumo2015} \\
Higgs measurements & $50 < m < 250$ & $< 0.09$ & TLEP \cite{Profumo2015} \\
\(q\bar{q} \to \phi W/Z\) \& \(pp \to t\bar{t}\phi\) & $m < 300$ & $< 0.31$ & LHC \cite{Arcadi2019} \\
\(\phi \to ZZ/WW\) & $m > 180$ & - & LHC \cite{Arcadi2019} \\
\(\phi \to t\bar{t}\) & $m > 350$ & - & LHC \cite{Arcadi2019} \\
Singlet direct production & $m > 1500$ & $< 0.48$ & CLIC \cite{deBlas2018, Franceschini2020} \\
\(\phi \to WW \to l\nu l\nu\) & $145 < m < 1000$ & - & ATLAS-CMS \cite{CMS2015} \\
\(\phi \to WW \to l\nu qq\) & $180 < m < 600$ & - & ATLAS-CMS \cite{CMS2015} \\
\(ZZ \to 2l2l'\) & $145 < m < 1000$ & - & ATLAS-CMS \cite{CMS2015} \\
\(ZZ \to 2l2\nu\) & $200 < m < 1000$ & - & ATLAS-CMS \cite{CMS2015} \\
\(ZZ \to 2l2q\) & $230 < m < 1000$ & - & ATLAS-CMS \cite{CMS2015} \\
\(pp \to \phi \to WW/ZZ\) & $m > 130$ & $< 0.37$ & LHC \cite{Ilnicka2018} \\
\hline
\end{tabular}
\label{tab:heavy}
\end{table}

\begin{table}[h]
\centering
\caption{Limits on the mixing angle and mass of the new scalar particle from various experiments for light mass ranges.}
\begin{tabular}{|l|c|c|l|}
\hline
\textbf{Experiment} & \textbf{\(m_{\phi}\) [GeV]} & \textbf{\(\cos \theta\)} & \textbf{Ref.} \\
\hline
BaBar (\(b \to s\phi\), \(\phi \to \mu^+\mu^-\)) & $< 5$ & $< 0.1$ & \cite{BaBar2013} \\
LEP (\(e^+e^- \to Z\phi\), \(\phi \to b\bar{b}, \tau^+\tau^-\)) & $< 40$ & $< 0.1-0.2$ & \cite{LEP2006, OPAL2003} \\
LHC Higgs coupling measurements & $< 50$ & $< 0.15-0.2$ & \cite{Robens2015} \\
\hline
\end{tabular}
\label{tab:light}
\end{table}

\newpage
\section{Theoretical Framework}

In this section, we consider the Standard Model with an extra singlet scalar boson \(\Phi\) which is symmetric under \(\Phi \rightarrow - \Phi\). Before proceeding, we clarify our notation to avoid confusion:

\begin{itemize}
    \item \(H\) denotes the SU(2)\(_L\) Higgs doublet field before electroweak symmetry breaking (EWSB)
    \item \(\Phi\) denotes the real singlet scalar field
    \item After EWSB, we expand around the vacuum expectation values:
    \[H = \frac{1}{\sqrt{2}}\begin{pmatrix} 0 \\ v_h + h' \end{pmatrix}, \quad \Phi = v_\phi + \varphi'\]
    \item The fields \(h'\) and \(\varphi'\) are not mass eigenstates; they mix via the mass matrix
    \item The physical mass eigenstates are obtained after diagonalization:
    \[\begin{pmatrix} \phi \\ h \end{pmatrix} = \begin{pmatrix} \cos\theta & -\sin\theta \\ \sin\theta & \cos\theta \end{pmatrix} \begin{pmatrix} h' \\ \varphi' \end{pmatrix}\]
    where \(h\) is the observed 125 GeV Higgs boson and \(\phi\) is the new scalar particle
\end{itemize}

In the Higgs portal model, the new scalar field couples to the SM fields only via the Higgs sector, which leads to a renormalizable term as in Eq. (\ref{eq:portal}), satisfying \(Z_2\) symmetry. The full Lagrangian for the Higgs sector including gauge interactions can be written as:

\begin{equation}
\mathcal{L} = \mathcal{L}_{\text{SM}} + \frac{1}{2}\partial_\mu\Phi\partial^\mu\Phi - V(\Phi,H) + \mathcal{L}_{\Phi\text{-gauge}},
\end{equation}

where \(\mathcal{L}_{\text{SM}}\) contains the Standard Model kinetic terms for the Higgs doublet including its covariant derivatives:

\begin{equation}
\mathcal{L}_{\text{SM}} \supset (D_\mu H)^\dagger (D^\mu H), \quad D_\mu = \partial_\mu - \frac{i}{2}g\tau^a W_\mu^a - \frac{i}{2}g' B_\mu.
\end{equation}
Although the singlet field $\Phi$ is a gauge singlet at tree level, mixing with the SM Higgs doublet induces couplings to electroweak gauge bosons. The physical mass eigenstates $h$ and $\phi$ couple to $W^\pm$ and $Z$ bosons with strengths proportional to $\sin\theta$ and $\cos\theta$, respectively. However, these gauge interactions do not contribute to the exotic decays $h \to \phi\phi$ or $h \to \phi\phi\phi$, which proceed solely through scalar self-interactions derived from the potential in Eq.~\eqref{eq:potential}. Consequently, while gauge couplings affect the total Higgs production cross-section and SM decay rates (which we account for via the $\sin^2\theta$ suppression), they have no direct impact on the partial widths computed in Section III.
The new scalar \(\Phi\) is a gauge singlet, so its gauge interactions \(\mathcal{L}_{\Phi\text{-gauge}}\) are zero at tree level. The scalar potential in unitary gauge, after expanding around the vacuum, can be written as:

\begin{equation}
V(\phi ,H) = \frac{1}{4}\lambda_{h}H^{4} + \frac{1}{4}\lambda_{\phi}\phi^{4} - \frac{1}{2}\mu_{h}^{2}H^{2} - \frac{1}{2}\mu_{\phi}^{2}\phi^{2} + \frac{1}{4}\lambda_{3}H^{2}\phi^{2}, \label{eq:potential}
\end{equation}

in which \(\mu_{h}\) and \(\mu_{\phi}\) are parameters with mass dimension, \(\lambda_{h}\), \(\lambda_{\phi}\) and \(\lambda_{3}\) are dimensionless coupling constants. Consequently, for \(\mu_{h}^{2} > 0\) and \(\mu_{\phi}^{2} > 0\), both \(H\) and \(\phi\) develop expectation values as follows:

\begin{equation}
v_{h}^{2} = \frac{4\lambda_{\phi}\mu_{h}^{2} - 2\lambda_{3}\mu_{\phi}^{2}}{4\lambda_{h}\lambda_{\phi} - \lambda_{3}^{2}}, \quad v_{\phi}^{2} = \frac{4\lambda_{h}\mu_{\phi}^{2} - 2\lambda_{3}\mu_{h}^{2}}{4\lambda_{h}\lambda_{\phi} - \lambda_{3}^{2}}, \label{eq:vevs}
\end{equation}

which leads to a non-diagonal mass matrix as

\begin{equation}
M_{h,\phi}^{2} = \left( \begin{array}{cc}\frac{\partial^{2}V(H,\Phi)}{\partial H^{2}} & \frac{\partial^{2}V(H,\Phi)}{\partial H\partial\Phi}\\ \frac{\partial^{2}V(H,\Phi)}{\partial\Phi\partial H} & \frac{\partial^{2}V(H,\Phi)}{\partial\Phi^{2}} \end{array} \right) = \left( \begin{array}{cc}2\lambda_{h}v_{h}^{2} & \lambda_{3}v_{h}v_{\phi}\\ \lambda_{3}v_{h}v_{\phi} & 2\lambda_{\phi}v_{\phi}^{2} \end{array} \right). \label{eq:massmatrix}
\end{equation}

This matrix can be transformed to the mass eigenstates by an orthogonal transformation:

\begin{equation}
\left( \begin{array}{c}\phi \\ h \end{array} \right) = \left( \begin{array}{cc}\cos \theta & -\sin \theta \\ \sin \theta & \cos \theta \end{array} \right)\left( \begin{array}{c}H \\ \Phi \end{array} \right), \label{eq:rotation}
\end{equation}

where \(\theta\) is the mixing angle and \(\cos \theta = 0\) is the SM limit. Subsequently, the eigenvalues and eigenstates of this new representation can be obtained as:

\begin{equation}
m_{h,\phi}^{2} = \lambda_{h}v_{h}^{2} + \lambda_{\phi}v_{\phi}^{2} \pm \sqrt{(\lambda_{h}v_{h}^{2} - \lambda_{\phi}v_{\phi}^{2})^{2} + (\lambda_{3}v_{h}v_{\phi})^{2}}, \label{eq:eigenvalues}
\end{equation}

\begin{equation}
\begin{array}{l}\phi = H\cos \theta -\Phi \sin \theta ,\\ h = H\sin \theta +\Phi \cos \theta . \end{array} \label{eq:eigenstates}
\end{equation}

Note that Eq. (\ref{eq:eigenvalues}) is the standard expression for the mass eigenvalues obtained from diagonalizing the mass matrix in Eq. (\ref{eq:massmatrix}) and matches, for example, Eq. (2.12) in Robens \& Stefaniak \cite{Robens2015}. The term inside the square root is correctly \((\lambda_h v_h^2 - \lambda_\phi v_\phi^2)^2 + (\lambda_3 v_h v_\phi)^2\), which follows from solving \(\det(M^2 - m^2 \mathrm{I}) = 0\).

Now one can consider \(h\) as the ordinary Higgs field in the SM with \(m_{h} = 125\) GeV and \(v_{h} = 246\) GeV and \(\phi\) can be defined as the new scalar field with mass \(m_{\phi}\). It is important to note that of the five parameters appearing in the scalar potential (\(\lambda_h, \lambda_\phi, \lambda_3, \mu_h, \mu_\phi\)), only three are independent. The two minimization conditions in Eq. (\ref{eq:vevs}) relate these parameters, leaving three free parameters that can be conveniently chosen as the physical masses \(m_h\), \(m_\phi\), and the mixing angle \(\theta\) (or equivalently, the VEVs \(v_h\) and \(v_\phi\)). The mixing angle \(\theta\) not only leads to new interactions in addition to the usual SM channels, but also corrects the coupling of the ordinary Higgs field in some interactions. The full set of Feynman rules for this extension of SM, including interactions with gauge bosons that arise from mixing-induced couplings, is given in Appendix A.

There are some relations that relate the parameters \((\lambda_{h},\lambda_{\phi}\) and \(\lambda_{3})\) to \((m_{\phi},v_{\phi}\) and \(\cos \theta)\) as follows:

\begin{equation}
\begin{array}{l}\lambda_{h} = \frac{m_{h}^{2}}{2v_{h}^{2}}\sin^{2}\theta +\frac{m_{\phi}^{2}}{2v_{h}^{2}}\cos^{2}\theta,\\ \lambda_{\phi} = \frac{m_{h}^{2}}{2v_{\phi}^{2}}\cos^{2}\theta +\frac{m_{\phi}^{2}}{2v_{\phi}^{2}}\sin^{2}\theta,\\ \lambda_{3} = \frac{m_{h}^{2} - m_{\phi}^{2}}{2v_{h}v_{\phi}}\sin 2\theta. \end{array} \label{eq:couplings}
\end{equation}

\section{Higgs decay rate to two and three real scalar particles}

In this section we examine the total decay rate of the SM Higgs particle to two and three real scalar particles to obtain the allowed range for the free parameters \(v_{\phi}\), \(\lambda_{3}\) and \(\lambda_{\phi}\). There is only one diagram at tree level for the two scalar particles decay. This decay rate by dimensional analysis is of the order of \(\lambda_{3}^{2}v_{h}^{2} / m_{h}\) where for \(\Gamma_{h\rightarrow 2\phi}\sim 0.03\) MeV the coupling should be of order of \(\lambda_{3}\sim 10^{-4}\). However, by using the appropriate vertex, the decay rate can be obtained as follows

\begin{equation}
\Gamma_{h\rightarrow 2\phi} = \frac{\mu^{\prime 2}}{8\pi m_{h}}\sqrt{1 - \frac{4m_{\phi}^{2}}{m_{h}^{2}}}, \label{eq:gamma2}
\end{equation}

where

\begin{equation}
\mu^{\prime} = -\frac{\sin 2\theta}{2v_{h}v_{\phi}} (\sin \theta v_{h} + \cos \theta v_{\phi})\left(m_{\phi}{}^{2} + \frac{m_{h}{}^{2}}{2}\right). \label{eq:mup}
\end{equation}

These expressions are in complete agreement with the well-established results in the literature, for example Robens \& Stefaniak \cite{Robens2015}, where the partial width for \(h \to \phi\phi\) is given (in their notation) by \(\Gamma(h \to ss) = g_{hss}^2/(32\pi m_h)\sqrt{1-4m_s^2/m_h^2}\) with the coupling derived from the scalar potential after diagonalization. Our \(\mu'\) corresponds to this coupling up to a factor of 2 (depending on convention for identical particles in the final state).

\begin{figure*}[h!]
		\begin{center}
			\begin{minipage}[b]{0.325\textwidth}\begin{center}
					\subfigure[]{
						\label{3decay1}\includegraphics[width=\textwidth]{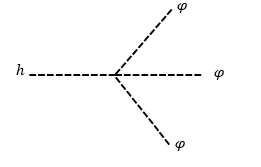}\qquad}
			\end{center}\end{minipage} \hskip+0cm 
			\begin{minipage}[b]{0.325\textwidth}\begin{center}
					\subfigure[]{
						\label{3decay2}\includegraphics[width=\textwidth]{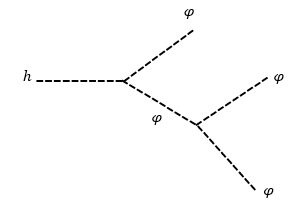}\qquad}
			\end{center}\end{minipage} \hskip0cm 
			\begin{minipage}[b]{0.325\textwidth}\begin{center}
					\subfigure[]{
						\label{3decay3}\includegraphics[width=\textwidth]{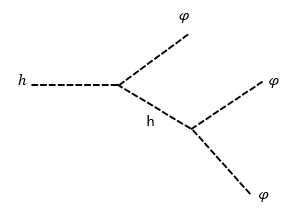}\qquad}
			\end{center}\end{minipage} \hskip0cm 
		\end{center}
		\caption{Feynman Diagrams of Higgs decay channels to three new scalar singlet. Diagram (a) is the contact diagram, while (b) and (c) involve intermediate scalar exchange and are suppressed in the small-mixing regime.}
		\label{3-decay}
	\end{figure*}

\begin{figure*}[h!]
		\begin{center}
			\begin{minipage}[b]{1.0\textwidth}\begin{center}
					\subfigure[]{
						\label{3hphi} \includegraphics[width=0.5\textwidth]{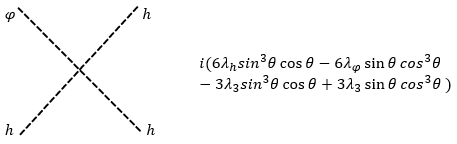}\qquad}
			\end{center}\end{minipage} \hskip+0cm 
			\begin{minipage}[b]{1.0\textwidth}\begin{center}
					\subfigure[]{
						\label{3phi}\includegraphics[width=0.5\textwidth]{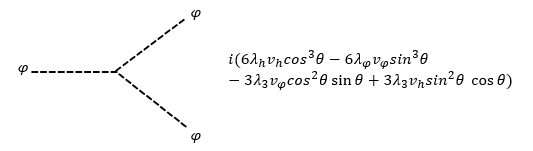}\qquad}
			\end{center}\end{minipage} \hskip0cm 
			\begin{minipage}[b]{1.0\textwidth}\begin{center}
					\subfigure[]{
						\label{h2phi}\includegraphics[width=0.5\textwidth]{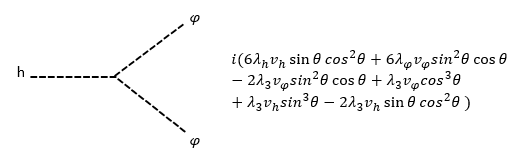}\qquad}
			\end{center}\end{minipage} \hskip0cm 
		\end{center}
		\caption{Feynman rules for Higgs-scalar particle couplings relevant for the Higgs decay channels to three new scalar singlet. The vertex factors are shown for each interaction.}
		\label{3-decay-vertex}
	\end{figure*}

Meanwhile, for the three scalar particles decay there are three different diagrams as shown in Fig. (\ref{3-decay}). To find the exact decay rate one should calculate all diagrams, which is very complicated, but for our aim of finding the order of magnitude of upper bounds on the parameters, a full calculation is not necessary. By comparing the amplitudes of these Feynman diagrams, one can find that only the first diagram is sufficient for our purpose. In fact, using the couplings for each diagram (Figs. \ref{3-decay-vertex} and \ref{fig:feynman_rules}) leads to decay rate ratios for the sub-figures a, b and c in Fig. (\ref{3-decay}), respectively, as \(\lambda^{2}\cos^{2}\theta :\lambda^{3}\cos \theta :\lambda^{4}\) for \(\cos \theta \ll 1\) and \(\sin \theta \simeq 1\). Furthermore, the diagrams involving intermediate \(h\) or \(\phi\) exchange are suppressed relative to the contact diagram by additional propagator denominators. Therefore, in the small-mixing regime \(\cos \theta \ll 1\), the contributions from non-contact diagrams are subleading by at least one to two orders of magnitude and one can only consider the contact term to calculate \(\Gamma_{h\rightarrow \phi \phi \phi}\) for \(\cos \theta >\lambda \sim 10^{-4}\).

\begin{figure}[h]
\centering
\includegraphics[width=0.6\textwidth]{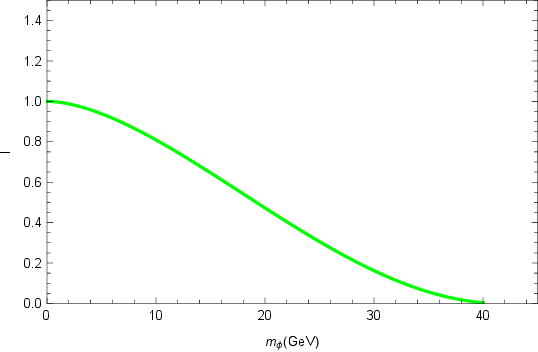}
\caption{The plot of \(I(m_{\phi})\) as a function of \(m_{\phi}\). The integral decreases as \(m_{\phi}\) increases due to phase space suppression.}
\label{fig:Im}
\end{figure}

To this end, one can obtain the decay rate as follows:

\begin{equation}
\Gamma_{h\rightarrow 3\phi} = \frac{81m_h}{64\pi^3}\frac{(\lambda_3 - 2\lambda_\phi)^2}{12}\sin^2\theta \cos^2\theta I(m_{\phi}), \label{eq:gamma3}
\end{equation}

where

\begin{equation}
\begin{array}{rl}
I(m_{\phi}) =& \int_{4m_{\phi}^2/m_h^2}^{(1-\sqrt{m_{\phi}^2/m_h^2})^2} (\frac{2}{s} (s^{4} + (\frac{-6m_{\phi}^{2}}{m_{h}^{2}} -2)s^{3} + (\frac{9m_{\phi}^{4}}{m_{h}^{4}} +\frac{6m_{\phi}^{2}}{m_{h}^{2}} +1)s^{2} \\
 +& (\frac{-4m_{\phi}^{6}}{m_{h}^{6}} +\frac{8m_{\phi}^{4}}{m_{h}^{4}} -\frac{4m_{\phi}^{2}}{m_{h}^{2}})s^{\frac{1}{2}}))ds, 
\end{array}\label{eq:I}
\end{equation}

and \(s = (p_{\phi_1} + p_{\phi_2})^2/m_h^2\) is the dimensionless invariant mass squared of two of the final state scalars. The integration limits correspond to the kinematic boundaries of the three-body decay: the lower limit \(4m_\phi^2/m_h^2\) occurs when two scalars are produced at rest relative to each other, and the upper limit \((1-\sqrt{m_\phi^2/m_h^2})^2\) occurs when one scalar takes almost all the available energy. The numerical values of \(I(m_{\phi})\) for different scalar masses are given in Fig. \ref{fig:Im}.


Therefore, adding (\ref{eq:gamma2}) and (\ref{eq:gamma3}) leads to the total decay rate as

\begin{equation}
\begin{array}{l}(\Gamma_{h\rightarrow 3\phi} + \Gamma_{h\rightarrow 2\phi})v_{\phi}^{4} = A\sin^{2}\theta \cos^{4}\theta v_{\phi}^{4}\\ \qquad +2Av_{h}\sin^{3}\theta \cos^{3}\theta v_{\phi}^{3}\\ \qquad +(Av_{h}^{2}\sin^{4}\theta \cos^{2}\theta +B\frac{(m_{h}^{2} - m_{\phi}^{2})^{2}}{v_{h}^{2}}\sin^{4}\theta \cos^{4}\theta)v_{\phi}^{2}\\ \qquad +( - 2B\frac{m_{h}^{2} - m_{\phi}^{2}}{v_{h}}(m_{\phi}^{2}\sin^{2}\theta +m_{h}^{2}\cos^{2}\theta)\sin^{3}\theta \cos^{3}\theta)v_{\phi}\\ \qquad +B(m_{\phi}^{2}\sin^{2}\theta +m_{h}^{2}\cos^{2}\theta)^{2}\sin^{2}\theta \cos^{2}\theta , \end{array} \label{eq:total}
\end{equation}

where

\begin{equation}
A(m_{\phi}) = \frac{10^{3}}{8\pi m_{h}}\sqrt{1 - \frac{4m_{\phi}^{2}}{m_{h}^{2}}}\left(m_{\phi}^{2} + \frac{m_{h}^{2}}{2}\right)^{2}\frac{1}{v_{h}^{2}}~(\mathrm{MeV}) \label{eq:A}
\end{equation}

\begin{equation}
B(m_{\phi}) = \frac{81m_{h}10^{3}}{768\pi^{3}} I(m_{\phi})~(\mathrm{MeV}), \label{eq:B}
\end{equation}

in which \(A\) and \(B\) are considered in \(\mathrm{MeV}\) units while \(m_{h}\), \(m_{\phi}\) and \(v_{h}\) are given in \(\mathrm{GeV}\) units.

The total Higgs decay width has been measured by the CMS collaboration using the off-shell production method. According to the Particle Data Group (PDG) 2022, the experimental value is \cite{PDG2022, CMS2019}:

\begin{equation}
\Gamma_h^{\text{exp}} = 3.2^{+2.8}_{-2.2}~\text{MeV}. \label{eq:width_exp}
\end{equation}

This value is consistent with the Standard Model prediction \(\Gamma_{\text{SM}}\) \cite{SMwidth} within the experimental uncertainties.

The dominant Standard Model decay channels (\(b\bar{b}\), \(WW^*\), \(gg\), \(\tau^+\tau^-\), \(c\bar{c}\), \(ZZ^*\)) account for more than 99\% of the total width \cite{ATLAS2016}. In our extension, the ordinary Higgs couplings are weakened by a factor \(\sin\theta\) due to mixing; therefore, all the Standard Model decay rates are reduced by \(\sin^2\theta\). Consequently, the modified total width from SM channels is \(\Gamma_{\text{NMD}} = \sin^2\theta\,\Gamma_{\text{SM}}\).

The remaining width available for exotic decays must satisfy:

\begin{equation}
\Gamma_{\text{NMD}} + \Gamma_{h\rightarrow 3\phi} + \Gamma_{h\rightarrow 2\phi} \lesssim \Gamma_h^{\text{exp}} \label{eq:constraint}
\end{equation}

Using the central value \(\Gamma_h^{\text{exp}} = 3.2\) MeV for our main analysis (and noting that the uncertainty does not qualitatively affect our bounds), we obtain:

\begin{equation}
\Gamma_{h\rightarrow 3\phi} + \Gamma_{h\rightarrow 2\phi} \lesssim 3.2(1 - 0.99\sin^2\theta)~\text{MeV}, \label{eq:bsm_bound}
\end{equation}

which leads to an equation for \(v_{\phi}\) as:

\begin{equation}
W_{4} = a_{4}v_{\phi}^{4} + a_{3}v_{\phi}^{3} + a_{2}v_{\phi}^{2} + a_{1}v_{\phi} + a_{0}\lesssim 0, \label{eq:quartic}
\end{equation}

where \(a_{4} - a_{0}\) are defined as follows

\begin{equation}
\begin{array}{rl} 
a_4 = & (A\sin^2\theta \cos^4\theta -3.2(1 - 0.99\sin^2\theta))~(\mathrm{MeV}),\\ 
a_3 = & 2Av_h\sin^3\theta \cos^3\theta,\\ 
a_2 = & \left(Av_h^2\sin^4\theta \cos^2\theta +B\frac{(m_h^2 - m_\phi^2)^2}{v_h^2}\sin^4\theta \cos^4\theta\right),\\ 
a_1 = & \left(-2B\frac{m_h^2 - m_\phi^2}{v_h}(m_\phi^2\sin^2\theta +m_h^2\cos^2\theta)\sin^3\theta \cos^3\theta\right),\\ 
a_0 = & B(m_\phi^2\sin^2\theta +m_h^2\cos^2\theta)^2\sin^2\theta \cos^2\theta . \end{array} \label{eq:coeffs}
\end{equation}

\subsection{Derivation of the bound on \(\cos\theta\)}

The bound on the mixing angle \(\cos\theta < 0.12-0.13\) arises from analyzing the quartic inequality in Eq. (\ref{eq:quartic}). The coefficient \(a_4\) in Eq. (\ref{eq:coeffs}) is particularly important:

\begin{equation}
a_4 = A\sin^2\theta \cos^4\theta - 3.2(1 - 0.99\sin^2\theta)~\text{MeV},
\end{equation}

where \(A(m_\phi)\) is given in Eq. (\ref{eq:A}) and is positive. For the inequality in Eq. (\ref{eq:quartic}) to have a physical solution with \(v_\phi > 0\), the quartic must open downward in the large \(v_\phi\) region, which requires \(a_4 < 0\). The condition \(a_4 < 0\) gives:

\begin{equation}
A\sin^2\theta \cos^4\theta < 3.2(1 - 0.99\sin^2\theta).
\end{equation}

For small mixing (\(\sin\theta \approx 1\), \(\cos\theta \ll 1\)), this simplifies to:

\begin{equation}
A\cos^4\theta \lesssim 3.2(1 - 0.99)=0.032.
\end{equation}

\begin{figure}[h]
\centering
\includegraphics[width=0.6\textwidth]{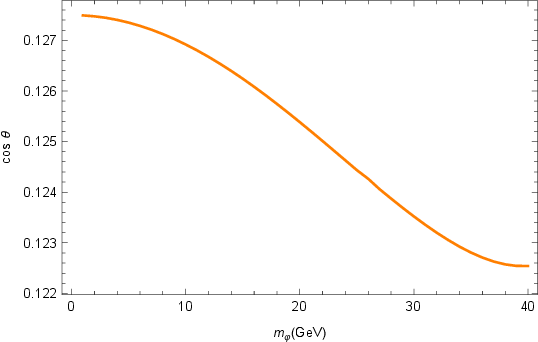}
\caption{The solid curve represents the value of \(\cos \theta\) for which the coefficient \(a_{4}\) in Eq. (25) vanishes, as a function of the scalar mass \(m_{\phi}\). The region below the curve corresponds to \(a_{4}< 0\), which is the physically allowed region. This yields the bound \(\cos\theta < 0.12-0.13\) across the mass range.}
\label{fig:a4}
\end{figure}

Since \(A(m_\phi)\) depends on \(m_\phi\) (through the phase space factor and the \((m_\phi^2 + m_h^2/2)^2\) term), solving \(a_4 = 0\) yields the critical curve shown in Fig. \ref{fig:a4}.

The solid curve in Fig. \ref{fig:a4} shows the value of \(\cos\theta\) for which \(a_4 = 0\) as a function of \(m_\phi\). The region \textbf{below} this curve corresponds to \(a_4 < 0\), which is the physically allowed region. For scalar masses in the range \(0 < m_\phi < 40\) GeV, this condition yields:

\begin{equation}
\cos\theta < 0.12 - 0.13
\end{equation}

with the mild \(m_\phi\) dependence giving the range \(0.13\) at low masses to \(0.12\) near \(m_\phi \sim 40\) GeV.

This bound is independent of \(v_\phi\) and represents a fundamental constraint from the requirement that the total exotic decay width does not exceed the available BSM width. It is complementary to direct search limits and provides an upper bound on the mixing angle that applies across the entire light scalar mass range.

As the sign of \(a_{4}\) is important for examining the inequality (\ref{eq:quartic}), we have solved \(a_{4} = 0\) for different \(m_{\phi}\) to find which range of \(\cos \theta\) leads to \(a_{4} < 0\); see Fig. \ref{fig:a4}.

However, for \(\cos \theta \ll 1\) one can easily show that \(a_{0}\) and \(a_{1}\) are very small, which leads to

\begin{equation}
v_{\phi}{}^{2}(a_{4}v_{\phi}{}^{2} + a_{3}v_{\phi} + a_{2})\lesssim 0. \label{eq:simplified}
\end{equation}

Therefore, to find the solution of the inequality (\ref{eq:simplified}) one should examine the sign of the quadratic equation \(a_{4}v_{\phi}{}^{2} + a_{3}v_{\phi} + a_{2}\) which has the following roots

\begin{equation}
v_{\phi}^{\pm} = \frac{a_{3}\mp\sqrt{a_{3}^{2} - 4a_{2}a_{4}}}{-2a_{4}}. \label{eq:roots}
\end{equation}

For \(\cos \theta \ll 1\) one can easily see that \(a_{4}< 0\), \(a_{2} > 0\) and \(a_{3} > 0\), and the inequality in Eq. (\ref{eq:quartic}) is satisfied for all values of \(v_{\phi}\) that lie outside the interval between the two roots given in Eq. (\ref{eq:roots}). Given that for these parameters \(v_{\phi}^{+}< 0\) and \(v_{\phi}^{-} > 0\), the allowed physical region (where \(v_{\phi} > 0\)) reduces to \(v_{\phi} > v_{\phi}^{-}\).

The vacuum expectation value $v_\phi$ is defined as $\langle\Phi\rangle$, the minimum of the effective scalar potential. For real $\Phi$ with $\mathbb{Z}_2$ symmetry ($\Phi \to -\Phi$), the potential is even in $\Phi$, and the minima occur at $\Phi = \pm v_\phi$. Choosing $v_\phi > 0$ is not merely conventional—it corresponds to selecting the global minimum of the potential where $\partial^2 V / \partial \Phi^2 > 0$, ensuring vacuum stability. A negative solution would describe the same physical vacuum due to the $\mathbb{Z}_2$ symmetry, so we restrict to $v_\phi > 0$ without loss of generality.

\begin{figure}[h]
\centering
\includegraphics[width=0.6\textwidth]{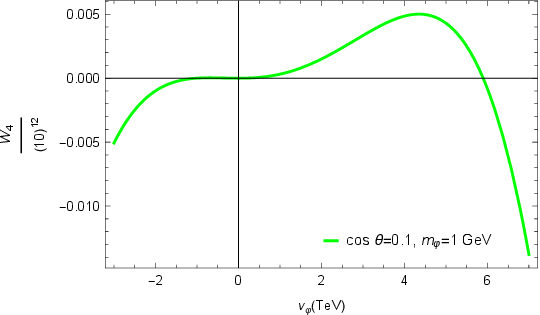}
\caption{The two roots, \(v_{\phi}^{+}\) and \(v_{\phi}^{-}\), of the quartic expression in Eq. (24) for \(\cos \theta = 0.1\) and \(m_{\phi}\simeq 1\) GeV. The inequality is satisfied for \(v_{\phi}< v_{\phi}^{+}\) or \(v_{\phi} > v_{\phi}^{-}\). Since \(v_{\phi}^{+}\) is negative and \(v_{\phi}\) is a physical vacuum expectation value (\(v_{\phi} > 0\)), the allowed parameter space is restricted to \(v_{\phi}\gtrsim 5.9\) TeV.}
\label{fig:vphi_root}
\end{figure}

For example, with \(\cos \theta = 0.1\) and \(m_{\phi}\simeq 1\) GeV, we find \(v_{\phi}^{+}< 0\) and \(v_{\phi}^{-}\simeq 6\) TeV. Consequently, the allowed range is \(v_{\phi}\gtrsim 6\) TeV. This result is confirmed numerically in Fig. \ref{fig:vphi_root}, which plots the roots of the quartic equation from (\ref{eq:quartic}) and shows agreement with the analytical expression (\ref{eq:simplified}). The figure indicates the allowed region begins at \(v_{\phi}\gtrsim 5.9\) TeV, consistent with the analytical calculation.

\begin{figure}[h!]
	\centering
	\begin{minipage}[b]{0.31\textwidth}
		\centering
		\includegraphics[width=\textwidth]{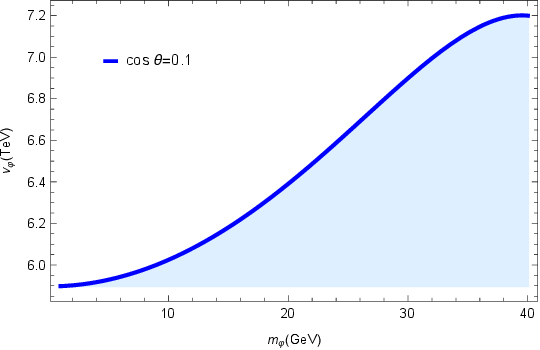}
		\label{fig:vphi}
	\end{minipage}
	\hfill
	\begin{minipage}[b]{0.325\textwidth}
		\centering
		\includegraphics[width=\textwidth]{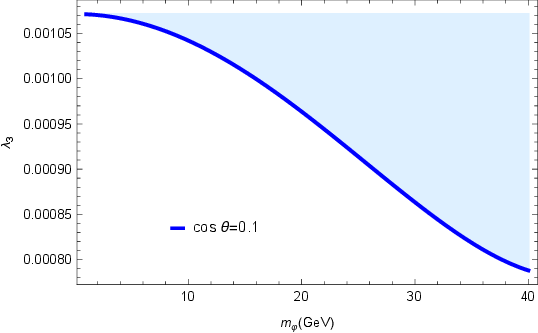}
		\label{fig:lambda3}
	\end{minipage}
	\hfill
	\begin{minipage}[b]{0.325\textwidth}
		\centering
		\includegraphics[width=\textwidth]{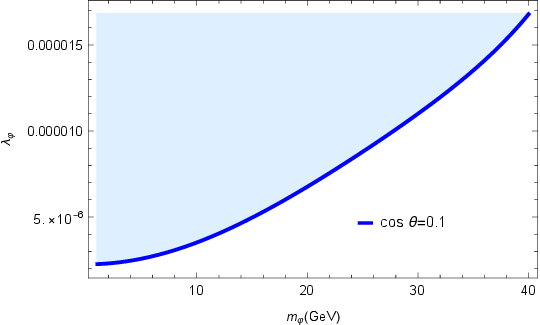}
		\label{fig:lambdaphi}
	\end{minipage}
	\caption{Values of \(v_{\phi}\) and the couplings \(\lambda_{3}\), and \(\lambda_{\phi}\) as functions of \(m_{\phi}\), for a fixed mixing angle \(\cos \theta = 0.1\). The curves are derived from the constraint that the sum of exotic Higgs decay rates saturates the allowed BSM width. As \(m_\phi\) increases, the minimum allowed \(v_\phi\) increases while \(\lambda_3\) decreases.}
	\label{fig:couplings}
\end{figure}

Furthermore, Fig. \ref{fig:couplings} shows the corresponding values of the couplings \(v_{\phi}\), \(\lambda_{3}\), and \(\lambda_{\phi}\) across the allowed scalar mass range. The curves in Fig. \ref{fig:couplings} are obtained through the following procedure:

\begin{enumerate}
    \item For a fixed mixing angle \(\cos\theta = 0.1\) and a given scalar mass \(m_\phi\), we solve the inequality in Eq. (\ref{eq:quartic}) to find the minimum allowed value of \(v_\phi\) that satisfies the total width constraint in Eq. (\ref{eq:bsm_bound}).
    
    \item This minimum \(v_\phi\) represents the smallest vacuum expectation value of the singlet field consistent with the Higgs width measurement for the chosen mixing angle.
    
    \item Using the relations in Eq. (\ref{eq:couplings}), we then compute the corresponding values of the couplings:
    \[\lambda_3 = \frac{m_h^2 - m_\phi^2}{2v_h v_\phi}\sin 2\theta,\]
    \[\lambda_\phi = \frac{m_h^2}{2v_\phi^2}\cos^2\theta + \frac{m_\phi^2}{2v_\phi^2}\sin^2\theta.\]
    
    \item The curves shown represent the maximum allowed values of these parameters (through the minimum \(v_\phi\)) consistent with the Higgs width constraint. For \(m_\phi\) approaching the kinematic threshold \(m_h/2\), the phase space suppression requires larger \(v_\phi\) (and correspondingly smaller \(\lambda_3\)) to satisfy the constraint.
\end{enumerate}

\subsection{Higgs decay rate to two real scalar particles}

\begin{figure}[h]
\centering
\includegraphics[width=0.6\textwidth]{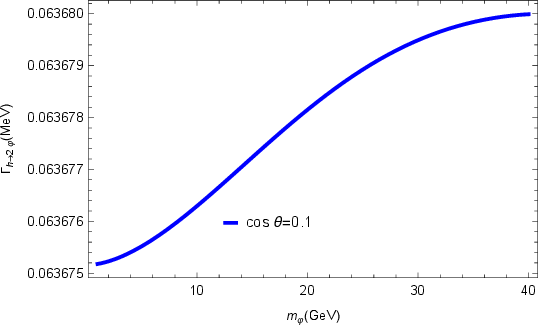}
\caption{The partial decay width \(\Gamma (h \to \phi \phi)\) as a function of the scalar mass \(m_{\phi}\), for a fixed mixing angle \(\cos \theta = 0.1\). The curve represents the maximum allowed rate under the constraint \(\Gamma (h \to \phi \phi) + \Gamma (h \to \phi \phi \phi) \leq \Gamma_{\mathrm{BSM}}\). The width remains approximately constant at 0.06 MeV due to the compensation between the decreasing phase space factor and the increasing coupling factors, with the squared coupling enhancing the compensation.}
\label{fig:2decay}
\end{figure}

Fig. \ref{fig:vphi_root} shows that the allowed parameter space for the vacuum expectation value is restricted to \(v_{\phi}\gtrsim 6\) TeV for \(\cos \theta = 0.1\). The corresponding values of \(v_{\phi}\) and the couplings \(\lambda_{3}\) and \(\lambda_{\phi}\), derived under this condition and the constraint \(\Gamma (h \to \phi \phi) + \Gamma (h \to \phi \phi \phi) \leq \Gamma_{\mathrm{BSM}}\), are plotted against \(m_{\phi}\) in Fig. \ref{fig:couplings}. Using these parameter values for different \(m_{\phi}\) in Eq. (\ref{eq:gamma2}), we calculate the Higgs decay rate to two singlet scalar particles, as shown in Fig. \ref{fig:2decay}.

The apparent constancy of \(\Gamma(h \to \phi\phi)\) as a function of \(m_\phi\) in Fig. \ref{fig:2decay} arises from a subtle compensation effect involving multiple factors in the coupling expression. The partial width is given by:

\[
\Gamma_{h\rightarrow 2\phi} = \frac{\mu^{\prime 2}}{8\pi m_{h}}\sqrt{1 - \frac{4m_{\phi}^{2}}{m_{h}^{2}}}
\]

where

\[
\mu^{\prime} = -\frac{\sin 2\theta}{2v_{h}v_{\phi}} (\sin \theta v_{h} + \cos \theta v_{\phi})\left(m_{\phi}{}^{2} + \frac{m_{h}{}^{2}}{2}\right).
\]

For fixed \(\cos\theta = 0.1\), as \(m_\phi\) increases:

\begin{enumerate}
    \item The phase space factor \(\sqrt{1-4m_\phi^2/m_h^2}\) decreases, which would normally reduce the partial width.
    
    \item The factor \(1/v_\phi\) in the coupling decreases because \(v_\phi\) increases with \(m_\phi\) (as shown in Fig. \ref{fig:couplings}), which also tends to decrease the partial width.
    
    \item However, the factor \((m_\phi^2 + m_h^2/2)\) in the coupling increases with \(m_\phi\), which increases the partial width.
    
    \item The factor \((\sin \theta v_h + \cos \theta v_\phi)\) also depends on \(v_\phi\), which increases with \(m_\phi\), providing an additional increase.
    
    \item The combination of the increasing factors \((m_\phi^2 + m_h^2/2)\) and \((\sin \theta v_h + \cos \theta v_\phi)\), together with the decreasing factors \(1/v_\phi\) and the phase space factor, results in a net compensation. Moreover, the coupling enters as \(\mu^{\prime 2}\) in the decay rate, so the compensation is even more effective, leading to a nearly constant value.
\end{enumerate}

The net effect is that \(\Gamma(h \to \phi\phi)\) varies by less than 1\% over the entire mass range 1-40 GeV, remaining at approximately 0.06 MeV. This represents the maximum allowed rate consistent with the Higgs width constraint.

\subsection{Higgs decay rate to three real scalar particles}

\begin{figure}[h]
\centering
\includegraphics[width=0.6\textwidth]{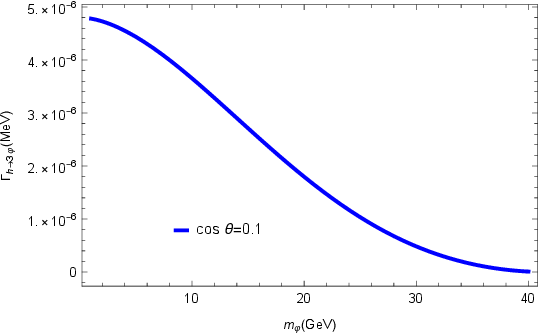}
\caption{The partial decay width \(\Gamma (h \rightarrow \phi \phi \phi)\) as a function of the scalar mass \(m_{\phi}\), for a fixed mixing angle \(\cos \theta = 0.1\). The curve represents the maximum allowed rate under the constraint \(\Gamma (h \rightarrow \phi \phi) + \Gamma (h \rightarrow \phi \phi \phi) \leq \Gamma_{\mathrm{BSM}}\). The three-body width is significantly smaller than the two-body width, reaching at most \(5 \times 10^{-6}\) MeV.}
\label{fig:3decay}
\end{figure}

The Higgs decay rate to three singlet scalar particles, \(\Gamma (H \to \phi \phi \phi)\), is also calculated as a function of the scalar mass \(m_{\phi}\) for a fixed mixing angle \(\cos \theta = 0.1\). This is achieved by substituting the values of \(v_{\phi}\), \(\lambda_{3}\), and \(\lambda_{\phi}\) obtained from Fig. \ref{fig:couplings} and dictated by the total width constraint into Eq. (\ref{eq:gamma3}). The resulting decay rate is presented in Fig. \ref{fig:3decay}.

\section{Conclusion}

In this work, we have performed a comprehensive analysis of the exotic Higgs decays within the framework of the Standard Model extended by a real singlet scalar field. While the two- and three-body channels themselves have been investigated previously, we used a consistent framework for evaluating the total exotic decay width \(\Gamma (h \to \phi \phi) + \Gamma (h \to \phi \phi \phi)\) to establish a constraint on the model's parameter space based on the measured total width of the Higgs boson.

We showed that the requirement \(\Gamma_{h \to \phi \phi} + \Gamma_{h \to \phi \phi \phi} < 3.2(1 - 0.99 \sin^{2} \theta)\) MeV leads to a fourth-order inequality in the singlet vacuum expectation value, imposing an upper limit on the Higgs-singlet mixing angle, \(\cos \theta < 0.12\), for \(m_{\phi} < 40\) GeV with a mild dependence on \(m_{\phi}(0.12 - 0.13)\).

Our bound of \(\cos \theta < 0.12\) provides a complementary and, in some regions, a leading constraint on the Higgs-singlet mixing. As summarized in Tables 1 and 2, existing limits vary significantly across the scalar mass range. For light scalars (\(m_{\phi} < 40\) GeV), our global constraint is particularly impactful. In regions where no direct limits exist, it provides the primary constraint. Where limits do exist---for instance, from LEP and LHC Higgs searches \cite{Robens2015}---our result is fully consistent, offering an independent and theoretically robust confirmation. Notably, for the mass range \(10 < m_{\phi} < 20\) GeV, our bound is competitive with the dedicated search limit of \(\cos \theta < \sim 0.11\) from Ref. \cite{Robens2015}. For heavier scalars (\(m_{\phi} > 50 \mathrm{GeV}\)), while dedicated collider searches often provide more stringent limits at specific high masses, our bound derived from the exotic decay width remains relevant as a universal, model-dependent constraint that applies across the entire viable mass spectrum below \(m_{h} / 2\).

When combined with existing external constraints that push the mixing angle closer to zero (e.g., \(\cos \theta < 0.1\)), our formalism allows for the precise determination of an upper bound on the singlet vacuum expectation value \(v_{\phi}\) and, consequently, on the individual exotic decay rates as \(\Gamma_{h \rightarrow \phi \phi} < 0.06 \mathrm{MeV}\) and \(\Gamma_{h \rightarrow \phi \phi \phi} < 5 \times 10^{-6} \mathrm{MeV}\).  The derived upper bound on the sum of exotic decay rates corresponds to a branching ratio well below the current LHC limit of $B(h\rightarrow invisible)<0.107$ \cite{atlas_cms_2023_invisible}, confirming the compatibility of our scenario with existing Higgs width measurements \cite{atlas2025}.

We note that our analysis uses the central value of the measured Higgs width. The experimental uncertainty (\(\pm 2.8\) MeV on the total width) would slightly relax or tighten the bounds, but does not change the qualitative features of our results. A more precise future measurement of the Higgs width would further sharpen the constraints presented here.

The methodology presented here---using the total exotic decay width as a global constraint---proves to be a valuable tool that is most powerful for light scalars where direct searches are challenging, while providing a crucial complementary bound across the entire low-mass range. To our knowledge, this is the first analysis to derive constraints on the Higgs-singlet mixing from the combined total exotic decay width, rather than from individual channels. This provides a genuinely global constraint on light-scalar portals.

\section*{Acknowledgments}

The authors would like to thank the referees for their constructive comments that helped improve this manuscript.

\newpage
\appendix
\section{Appendix: Feynman Rules}

\begin{figure}[h]
\centering
\includegraphics[width=0.9\textwidth]{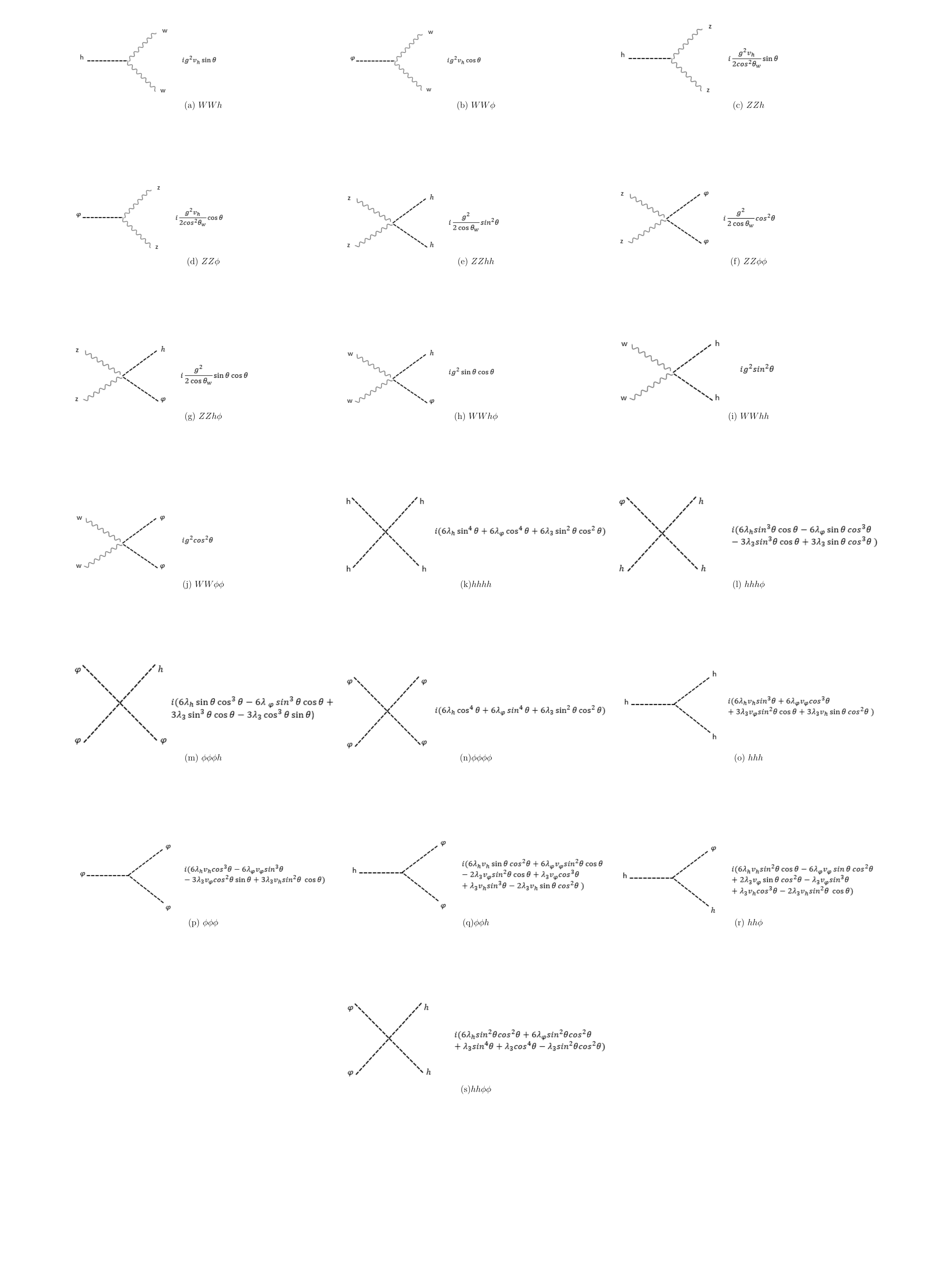}
\caption{Complete set of Feynman rules for the Standard Model extended by a real singlet scalar field \(\phi\). All vertex factors are displayed with their corresponding diagrams. The interactions with gauge bosons (hVV, hhVV, etc.) are modified by factors of \(\sin\theta\) and \(\cos\theta\) relative to the SM couplings.}
\label{fig:feynman_rules}
\end{figure}

The complete set of Feynman rules for the Standard Model extended by a real singlet scalar field \(\phi\) is presented in Fig. \ref{fig:feynman_rules}. All vertex factors are calculated in the convention where all momenta are considered incoming. These rules include interactions with gauge bosons, which arise from the mixing-induced couplings between the physical Higgs \(h\) and the gauge bosons, as well as self-interactions of the scalar fields.

\end{document}